# Data-driven analysis of the $\gamma\gamma^* \to \pi^0$ system using mathematical models and the role of feedback-loop dynamics

N. G. Stefanis[1,*]

[1] *Institut für Theoretische Physik II, Ruhr-Universität Bochum, D-44780 Bochum, Germany*

The data behavior of the pion-photon transition form factor (TFF) is discussed using a nonlinear mathematical model with two parameters $B$ and $C$. Analysis shows that the model's inherent inhibition provides a precondition for the asymptotic saturation of the TFF in agreement with perturbative QCD. Integral to this derivation is the use of a novel fundamental relation between the half-saturated TFF and its asymptotic limit given by $B$. This helps avoid the determination of the location of the TFF at infinite momentum by estimating instead the maximum slope of the curve at $Q^2 = C$. In conjunction with the $1\sigma(B, C)$ confidence ellipse, this framework provides an improved fit to the Belle data by setting stronger constraints on the fitting parameters and taking into account the sensitivity to third-party events below Belle's kinematic coverage. Without inhibition, the slope of the TFF curve would continue to grow so that saturation towards the asymptotic QCD limit would not be achieved. In order to compare the key features of TFF fit models to upcoming high-precision data in a standardized way, a conformity protocol in terms of QCD-based criteria is worked out. Adopting a broader perspective, we show that the asymptotic saturation of an inhibited TFF fit curve is analogous to a mechanical system driven by a feedback-loop mechanism in control theory.

*Introduction.* The neutral pseudoscalar $\pi^0$ meson is one of the three lightest hadrons $\pi^+, \pi^-, \pi^0$ consisting of a valence quark-antiquark pair bound by strong interactions described by quantum chromodynamics (QCD). Therefore, understanding its structure in terms of the twist-two (tw-2) light-cone distribution amplitude (DA) (which is the pion wave function integrated over transverse momenta) is of paramount importance. The pion DA is a universal, though not directly observable, pion characteristic that enters as the main nonperturbative input various hard exclusive processes in the context of collinear factorization within perturbative QCD [1].

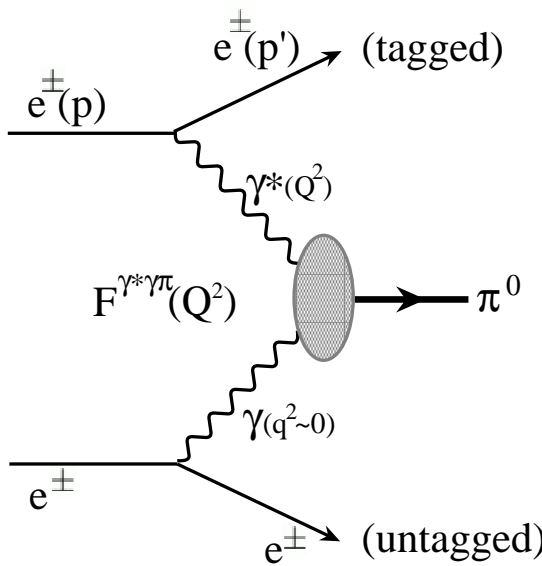

FIG. 1: Schematic Feynman diagram for the process $e^{\pm}e^{\pm} \to e^{\pm}e^{\pm}\pi^0$ involving the transition form factor $F^{\gamma^*\gamma\pi^0}(Q^2)$ to describe the two-photon process $\gamma^*\gamma \to \pi^0$ in a single-tag experimental mode. The shaded oval represents the pion DA.

For instance, though the Coulomb form factor of $\pi^0$ vanishes, the properties of its electromagnetic vertex can be revealed in single-tagged $e^+e^-$ experiments by measuring the momentum dependence of the transition form factor (TFF) $F^{\gamma^*\gamma^*\pi^0}\left(q_1^2, q_2^2\right)$, which describes the process $\pi^0 \to \gamma^*\gamma^*$ in the spacelike region. To this end, one employs a signal kinematics to select events in which the $\pi^0$ meson and one final-state electron (or positron)—the "e(p)-tag"—are registered, while the other lepton remains undetected because it is scattered at a very small angle. In this case, the virtual photon emitted from the tag has a large virtuality $Q^2 \equiv -q_1^2 = (p - p')^2$, where $p$ and $p'$ are the four momenta of the initial/final leptons. The other photon has a very low virtuality $Q^2 \gg q^2 \equiv -q_2^2 \gtrsim 0$ because the momentum transfer to the untagged "electron", from which it is virtually emitted, is close to zero (see Fig. 1). The measurements of the $\pi^0$ production via a two-photon process in single-tag experiments can be expressed by the scaled TFF

$$\mathcal{F}(Q^2) \equiv Q^2 F^{\gamma^*\gamma\pi^0}(Q^2) \tag{1}$$

and is defined in relation to of the matrix elements

$$\int d^4z\, e^{-iq_1\cdot z}\langle\pi^0(P)|T\{j_\mu(z)j_\nu(0)\}|0\rangle = i\epsilon_{\mu\nu\alpha\beta}q_1^\alpha q_2^\beta \times\ F^{\gamma^*\gamma^*\pi^0}(Q^2, q^2)\,, \tag{2}$$

where $j_\mu = \frac{2}{3}\bar{u}\gamma_\mu u - \frac{1}{3}\bar{d}\gamma_\mu d$ is the quark electromagnetic current. It represents the deviation of the meson production rate of the tag in comparison to point-like mesons and describes the effect of strong interactions on the electromagnetic $\gamma^*\gamma\pi^0$ transition amplitude. In leading order of perturbative (p)QCD, the TFF is given by the convolution [1, 2]

$$F^{\gamma^*\gamma\pi^0}(Q^2) = N_{\rm T}\int_0^1 dx T(Q^2, \mu_{\rm F}^2, x)\varphi_\pi^{(\text{tw-2})}(x, \mu_{\rm F}^2)\,, \tag{3}$$

where $T$ represents the leading-power term $\sim 1/Q^2$, at the factorization scale $\mu_{\rm F}$ with $N_{\rm T} = \sqrt{2}f_\pi/3$, where $f_\pi \approx 132$ MeV is the decay constant of the pion determined from leptonic decays [3]. From the experimental side, single-tag measurements have been reported in [4–7]. More recently, the BESIII Collaboration released preliminary data with an unprecedented accuracy measured at the BEPCII collider [8, 9] in the range $(0.31 \leq Q^2 \leq 2.95)$ GeV$^2$ (see Fig. 2).

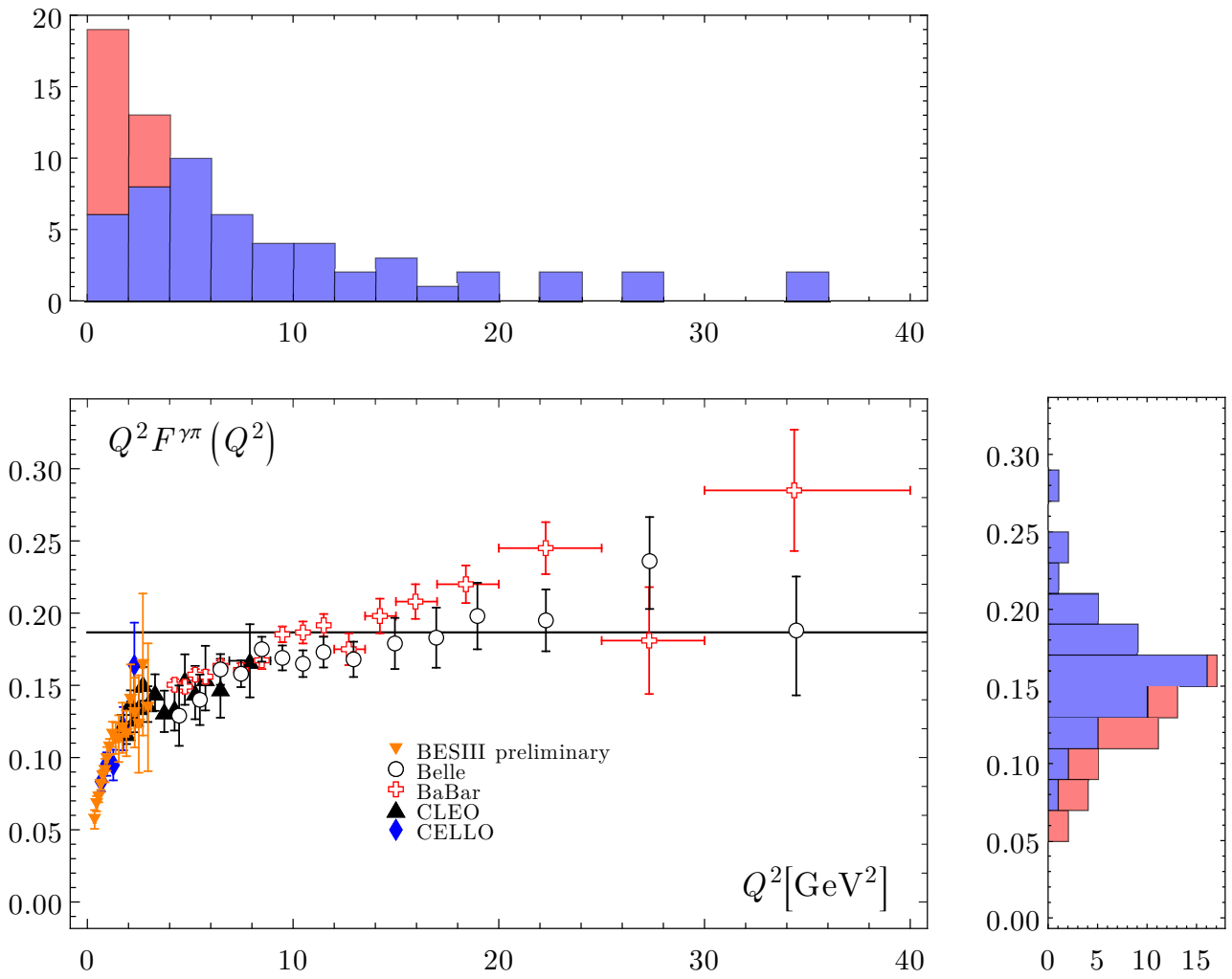

FIG. 2: Center: Pion-photon TFF data from single-tag experiments [4–8] with labels given inside. The error bars for $Q^2F^{\gamma\pi}(Q^2)$ are the sum in quadrature of statistical and systematic uncertainties. The solid horizontal line corresponds to the asymptotic pQCD limit given in Eq. (6). Top: Histogram showing the spread of the data over the momentum range $Q^2 \in [0, 40]$ GeV$^2$ in 20 bins with a width of 2 GeV$^2$. Right: Histogram displaying the frequency of occurrence of the measured $Q^2F^{\gamma\pi}(Q^2)$ GeV values over a range of 15 bins in the interval $[0.01-0.31]$ GeV with an increment of 0.02 GeV. Bins of BESIII data are shown in red color on top of the others.

*Scope of the work.* In this work we do not employ modeling of the pion DA to derive pion-photon TFF predictions within particular theoretical schemes, though we keep in mind the QCD background presented in the Supplemental Material and elsewhere [10]. We perform instead a data-driven analysis using results for the TFF that support inhibition [7]. We also consider the characteristics of fitting curves which indicate unrestricted growth at large $Q^2$.

*Mathematical TFF Model.* To be specific, we create for the TFF growth curve a mathematical "twin" of the Belle data in terms of a mathematical model, termed Fit(B) [7], which has the following functional structure

$$\mathcal{F}(Q^2) = \frac{BQ^2}{C + Q^2} \,. \tag{4}$$

The dependent variable $\mathcal{F}$, defined in Eq. (1), describes the production rate of the tag with respect to the parameters $Q^2 = 2E_{\rm beam}E_{\rm tag}\,(1 - \cos\theta_{\rm tag})$, where $E_{\rm beam}$ and $E_{\rm tag}$ are the energies of the tag before and after scattering, while $\theta_{\rm tag}$ is the scattering angle of the tag. Because $\mathcal{F}$ cannot be expressed as a linear combination of $B$ and $C$, no closed-form expression between the best-fitting parameters can be obtained as in linear regression. Thus, the production rate of the tag depends on the relative size of the parameters $Q^2/C$ and $B$ as $Q^2$ varies. This nonlinear variation can be expressed with respect to the $1\sigma$ confidence ellipse in the $(B, C)$ plane see Fig. 6 in [11]. This way, the two model parameters $B$ and $C$ can be determined empirically to reproduce the behavior of the measured TFF growth curve that maps a large part of the essential QCD processes. In particular, Fit(B) is sufficient to reproduce the steep growth of the TFF at small $Q^2$ values, where the single-tagged production rate grows almost linearly for a large amount of measurements (see top panel of Fig. 2). On the other hand, further increase of $Q^2$ to larger values would not entail further growth of the production rate of the tag because this has already reached the constant saturation limit $B$ at $Q^2 \to \infty$. Remarkably, this behavior of Fit(B) avoids interventional model elements, like a pion DA, to capture the right TFF growth behavior expected from QCD.

*Key elements of Fit(B).* As benchmark $B$ and $C$ parameters we use the fit results obtained by the Belle Collaboration [7]

$$\begin{aligned} B &= 0.209 \pm 0.016 \text{ GeV} \\ C &= 2.2 \pm 0.8 \text{ GeV}^2 \end{aligned} \tag{5}$$

with the goodness of fit $\chi^2/\text{ndf} = 7.07/13$. An independent data fit [11] yields numerically similar results so that mathematical projection and observation go well together. The reason is that Fit(B) lends itself by construction to the saturation behavior of the TFF in the pQCD asymptotic limit:

$$\mathcal{F}_\infty \equiv \lim_{Q^2\to\infty} Q^2F(Q^2) = \sqrt{2}f_\pi \sim 0.187 \text{ GeV} \,. \tag{6}$$

Ultimately, as evidenced by detailed analysis further down in the text, the stability of the mathematical model for the TFF growth curve in the asymptotic region can be fathomed in analogy to a self-stabilized system due to a feedback loop mechanism [12] which creates controlled inhibition.

*Halfway-saturated TFF.* While $B$ represents the maximum TFF in the asymptotic limit, on par with the maximum production rate of the tag, leading to saturation, $C$ is the amount of the TFF needed to reach half of $\mathcal{F}_{\rm max}$. This is actually the defining feature of $C$ and is reached when the momentum $Q^2$ becomes numerically equal to the parameter $C$. Then, the growth curve of the TFF reaches its maximum slope, see left panel of Fig. 3. As a result, one obtains the following *exact* relation

$$\mathcal{F}_{1/2}(Q^2 = C) = \frac{B}{2} \,. \tag{7}$$

This novel bridging relation between the low-$Q^2$ domain of the TFF and its asymptotic regime provides a stringent constraint on Fit(B). Moreover, it imposes a calibration condition on $B$ at the *finite* value $\log(Q^2/C)$ and avoids the estimation of $\mathcal{F}_max = B$ at infinity (see left panel of Fig. 3) using instead $\log Q^2/C$ because just a few logarithmic units suffice to estimate the maximum slope. For the Belle fit values (5) it yields $B/2 = (0.1045 \pm 0.008)$ GeV at the momentum scale $C = (2.2 \pm 0.8)$ GeV$^2$. These values correspond to a location well below the lowest Belle data point at $Q^2 < 4.46$ GeV$^2$ [7] (see Fig. 2). Thus, (7) can be used to test the compatibility of the Belle data with third-party events below the data range covered by the Belle measurements and eventually use them to improve the Belle fit. This synthetic data fitting procedure (constrained by the $1\sigma(B, C)$ confidence ellipse (see Fig. 6 in [11]) is carried out below and the results are shown in the right panel of Fig. 3.

*Segmentation of Fit(B) vs $Q^2$.* To disentangle the dynamical behavior of Fit(B) at different momentum scales, we perform a segmentation of its growth curve in relation to phases characterized by a distinct $[Q^2]^n$ dependence and monitor its development for $n = 0, 1, 2$, see Table I. Referring to this table, we observe that the first segment S1 contains the burst phase of the TFF which can be approximated by a steep linear growth $\sim [Q^2]^1$, so that the production rate of the tag has a first-order ($n = 1$) $Q^2$ dependence. This behavior reflects the hadronic shadow of the real photon in the QCD description of the TFF [10] in this regime and is illustrated in the right panel of Fig. 3. It shows the end of the initial linearly growing phase of the TFF at the half-saturation point, see Eq. (7), and is characterized by the maximum slope at an angle of 30° shown in the left panel. It reflects the first-order ($n = 1$) behavior of the TFF in segment S1.

In S2 the TFF shows a mixed-order metastable behavior, which is controlled by the nonlinear equation (4) for $Q^2 \geq C$. In the context of QCD, this complex behavior may be attributed to a mixture of nonperturbative contributions (higher twists) and pQCD radiative corrections beyond leading order that enter the TFF with different signs depending on $Q^2$ [10, 13, 14]. The asymptotic regime of the TFF is contained in S3. Because $Q^2 \gg C$, Fit(B) in S3 can be approximated by $\mathcal{F} = \mathcal{F}_{\max}$. In the asymptotic limit, this equals $B$ so that the TFF becomes a constant showing zero-order behavior. This is related to the fact that the tag production rate ceases to increase and becomes completely saturated. It further implies that the two quarks in the $\pi\gamma\gamma^*$ system can be described by a DA evolved to the asymptotic limit $\varphi_\pi^{\text{asy}}(x) = 6x\bar{x}$ [15] without any further involvement of hadron binding.

*Slope behavior of TFF.* The above discussion is based on the role of the slope behavior of the TFF curve and is shown in the left panel of Fig. 3. Indeed, the slope provides a measure of the sensitivity of the TFF to the intrinsic inhibition of the tag production rate. In this respect, the dynamical behavior of the inhibited production rate of the tag is akin to the widely used Michaelis-Menten equation to analyze the kinetics of enzyme-catalyzed reactions which show saturation at high substrate concentration [16, 17]. Crucially, the slope has a maximum value with a tangent at about 30° which marks the turning point in S1 from the steep production phase of the tag to the saturation regime in S3, where the TFF levels off (left panel of Fig. 3). As one sees from this figure, the tangent line is sufficiently straight over two logarithmic units in S2, assuming that abscissa and ordinate are measured in the same units. Without inhibition, the slope would continue to grow in S2 above $Q^2 > C$ so that the TFF would never enter the saturation regime predicted by QCD at high $Q^2$. Prioritizing the QCD asymptotic limit, one should navigate these complex issues carefully because to a certain extent the underlying explanation of uninhibited slope (TFF) behavior would be a competitor to Eq. (6) [2, 11, 18, 19].

TABLE I: Dynamical segmentation of the "twin" growth curve of the TFF fitted to the Belle data with Fit(B) to replicate its behavior. This is given by Eq. (4) with respect to the squared momentum $Q^2$ transferred by the photon emitted by the tag. Uninhibited linear growth in the initial rate period in S1 is followed by moderate mixed-order growing behavior characterized by metastability in S2. Saturation occurs in S3 when the TFF approaches asymptotically its maximum value given by the constant $B$ at $Q^2 \to \infty$.

| Segment | TFF | Process order | Phase |
|---|---|---|---|
| S1: $Q^2 \ll C$ | $\mathcal{F} = BQ^2/C$ | first order | uninhibited |
| S2: $Q^2 \geq C$ | $\mathcal{F} = BQ^2/(C + Q^2)$ | mixed order | metastable |
| S3: $Q^2 \gg C$ | $\mathcal{F}_{Q^2\to\infty} = B$ | zero order | saturated |

*Belle vs exogenous data.* Visual inspection of figure Fig. 2 shows that the Belle best-fit parameters given in (7) do not match with the topology of the exogenous data presented graphically in the center panel of Fig. 2. Therefore, in order to improve the compatibility of the fitted TFF with these data, we propose to determine refined values of the parameters $B$ and $C$ making use of their $1\sigma$ confidence ellipse worked out in [11] and shown in Fig. 6 there. This is done in conjunction with the novel half-saturation relation (7) which takes care that the improved $B, C$ parameters satisfy the appropriate calibration condition.

*Synthetic fitting procedure.* Following this strategy, we select best-choice $B, C$ values in the near-end region of the major axis of the $1\sigma(B, C)$ confidence ellipse [11] to determine $B \in [0.190 - 0.194]$ GeV and $C \in [1.2 - 1.4]$ GeV$^2$. Using the central values of these rather prudent intervals, we obtain with Eq. (7)

$$\begin{aligned} B^* &= 0.192 \text{ GeV} \\ C^* &= 1.3 \text{ GeV}^2 \,. \end{aligned} \tag{8}$$

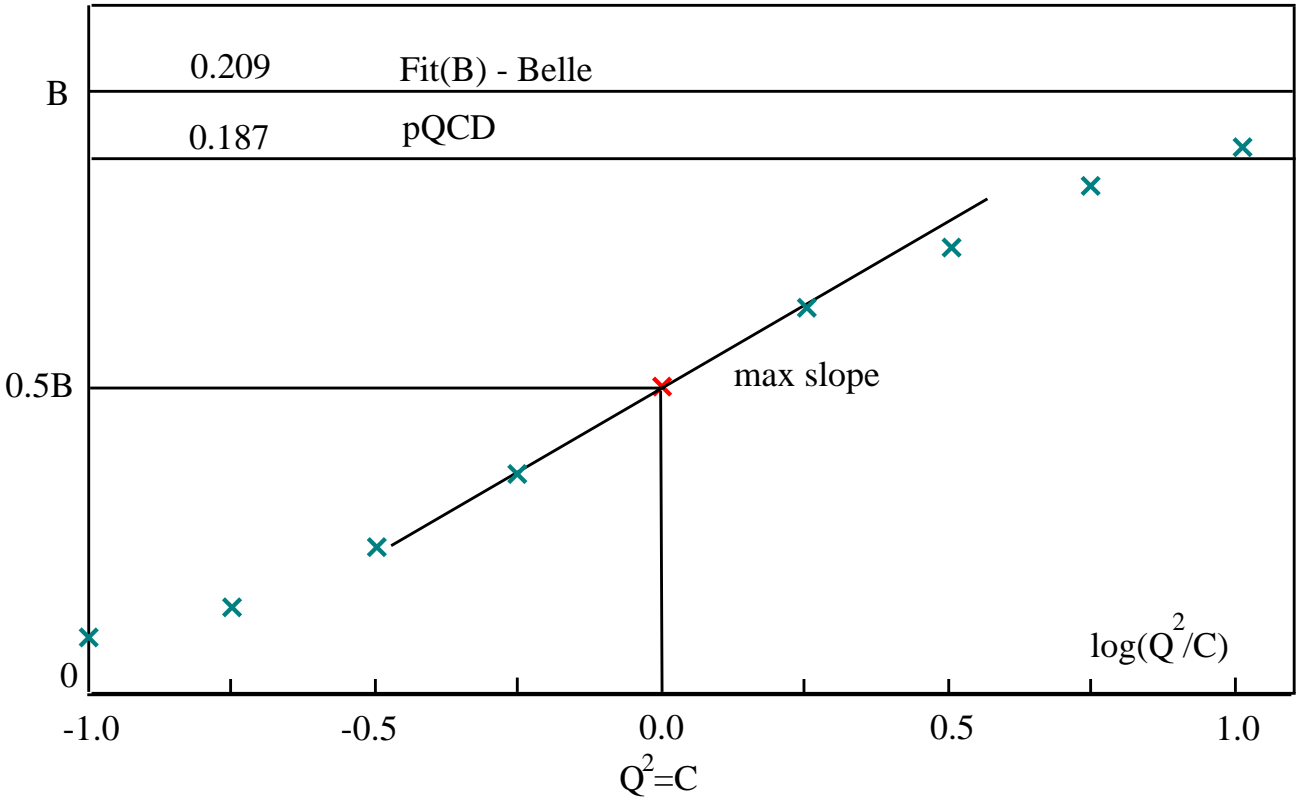

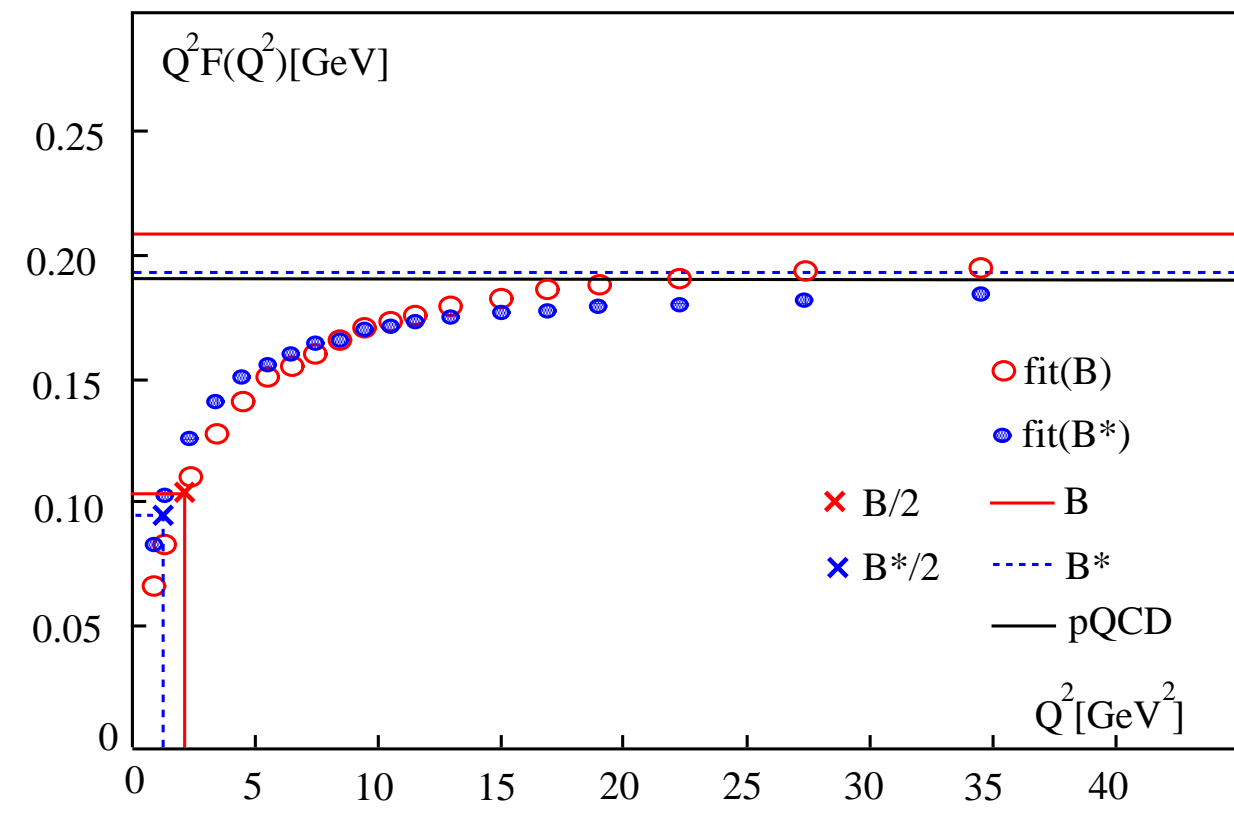

FIG. 3: Left: Semi-logarithmic plot of the TFF growth curve using Fit(B) of the Belle data [7]. The burst phase in S1 shows linear increase characterized by the maximum slope $B/C$ at $Q^2 = C$ with a tangent angle of 30°. The red cross marks the begin of inhibition in S2 when the TFF is half of $B$. Right: Curves of Fit(B), given by Eq. (4), for the Belle data using for $B$ and $C$ the central values of the Belle estimates [7] (open red circles). The improved values (8) are shown as filled blue circles. The corresponding positions of the half-saturated TFF values are indicated by crossing perpendicular lines. Points in the range $Q^2 \in [4.46 - 34.46]$ GeV$^2$ represent Belle events, while points at lower $Q^2$ values serve for illustration of the fitted TFF curves.

These values agree well with the lower limits of the corresponding Belle estimates of the $B$, $C$ parameters given by (5) and are also within the $1\sigma$ confidence region [11]. Then, the half-saturated TFF has the value $B^*/2 = 0.096$ GeV at $C^* = 1.3$ GeV$^2$ which is close to the CELLO event $0.0954^{+0.01}_{-0.0112}$ GeV in the interval [1.1−1.5] GeV$^2$—see Table III in [10]. Also the data point $0.116 \pm 0.009$ GeV of BESIII at $Q^2 = 1.226$ GeV$^2$ (see [20]) appears to be in the neighborhood of the improved $B$ parameter. This optimized Fit(B) curve is graphed in the right panel of Fig. 3 with the help of filled blue points in comparison with the original Belle fit (open red points). The convergence of the physical (i.e., data) world and the mathematical world of the improved twin model provides a better measure for the long trend of the data compared to the original Fit(B) [7] .

*Uninhibited TFF behavior.* The key element in the above considerations with respect to the analysis of the Belle data using Fit(B), is its inherent inhibition demonstrated by the slope behavior of the TFF curve. If there is no inhibition, as indicated by the pattern of growing TFF values measured by *BABAR* above $\sim 10$ GeV$^2$ (see Fig. 2), then it would be better suited to use the following parametrization (termed Fit(A) [6]),

$$\mathcal{F}(Q^2) = A \left( \frac{Q^2}{10 \mathrm{GeV}^2} \right)^{\beta} , \tag{9}$$

where $A$ and $\beta$ are fit parameters. Their values can be found in [6, 7], while the associated $1\sigma$ error ellipse was determined in the right panel of Fig. 6 in [11]. To make our point here, it is more important to concentrate on the calibration coefficient 10 GeV$^2$ in the denominator of Eq. (9) because just this scale marks the crossing point of the fitting curve with the asymptotic TFF line. It is obvious that at this momentum value, the parenthesis reduces to unity and the TFF becomes equal to the fitted parameter $A = 0.182$ GeV (omitting uncertainties). In fact, the adjustment of the calibration scale to 10 GeV$^2$ helps fixing the location of the crossing point closer to the QCD asymptotic line at 0.187 GeV (see Fig. 2). Above this crossing point, the fitted TFF continues to grow with the power $\beta = 0.25 \pm 0.02$ and is unable to reach a final state of saturation (even above the QCD limit) because there is no inhibition of the TFF growth curve. In other words, the slope of the TFF is prevented from reaching a maximum as in the case of Fit(B) (see left panel of Fig. 3). Therefore, the TFF has no zero-order phase asymptotically.

*Origin of inhibition.* We now proceed to give a more fundamental explanation of the intrinsic damping mechanism underlying the inhibited behavior of the TFF in Fit(B) accompanied by the appearance of a zero-order TFF phase asymptotically. Performing a simple rearrangement of expression (4), we show that it can be expressed as a backward mapping of the fitted asymptotic parameter $B$ to any earlier value of the TFF to give

$$Q^2 F(Q^2) = \frac{Q^2}{1 + Q^2/C} B/C \, , \tag{10}$$

where the mapping operator $R$ is produced by the feedback mechanism:

$$R = \frac{Q^2}{1 + Q^2/C} \, . \tag{11}$$

To expose the connection to a feedback-loop controlled system for a motor operator with negative feedback, the terminology used by Wiener in [12] is adopted. Consider a generic mechanical system with a feedback mechanism. Then, the motor operator with negative feedback is given by (11), with $Q^2 = A$ and the feedback operator reads $R = A/(1 + \lambda A)$ , where the multiplier operator is $\lambda = 1/C$ and $X = B/C$. The behavior of this system is shown in Fig. 4 and represents the flow of $\mathcal{F}(Q^2)$ [GeV] in terms of $AY$. The key element is the feedback operator (11) which modulates the power of the motor by the multiplier $\lambda$ in such a way as to keep the system on a stable dynamical phase. This resembles the inhibited behavior of the TFF in the zero-order region S3, where a steady state characterized by the constant $B$ is approached asymptotically.

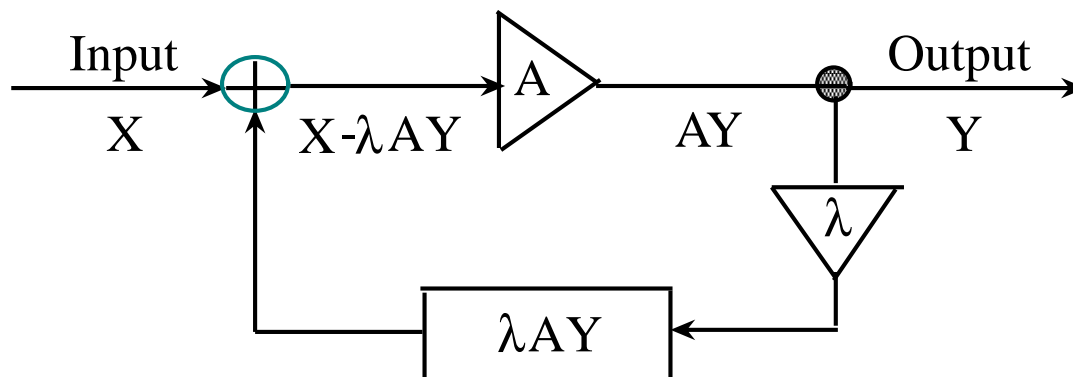

FIG. 4: Control flow chart for a generic mechanical system with an idealized negative feedback mechanism. It contains the following basic elements: (i) Original Input $X$, (ii) Subtractor $\bigoplus$, (iii) Motor Input $Y = X - \lambda AY$ after subtracting from the original input the output of the multiplier operator.

*Conformity protocol.* Here we formulate a conformity protocol of the fit functions Fit(A) and Fit(B) with respect to criteria developed and used in this analysis. This protocol is given in Table II. It can be used as a standardized test of different data sets or model calculations of the TFF expressed with reference to these fit models fared against QCD-based criteria. For real data fits, we refer to [6, 7, 11].

TABLE II: Conformity protocol of the growth curve of the TFF fitted by Fit(A), (9), or Fit(B), (4), against various criteria derived in this work. The ticks (Yes) and crosses (No) indicate how well each fit fares in relation to each criterion.

| Criterion | Fit(A)$(A, \beta)$ | Fit(B)$(B, C)$ |
|---|---|---|
| Calibration | crossing $\mathcal{F}_\infty$ | $\mathcal{F}_{1/2} = B/2$ |
| Best fit $\chi^2$ | ✓ | ✓ |
| $1\sigma$ error ellipse | ✓ | ✓ |
| Inhibition | ✗ | ✓ |
| Saturation | ✗ | ✓ |
| pQCD limit | ✗ | $B$ |
| $\mathcal{F}$ slope | grows | max at $Q^2 = C$ |
| Feedback loop | ✗ | ✓ |

*Summary and conclusions.* In this work we addressed the data for the pion-photon TFF in terms of two mathematical models each containing two fit parameters.

We identified the calibration conditions for both fits and investigated in detail the far-end behavior of the TFF with respect to $Q^2$. We showed that the intrinsic inhibition of Fit(B) inevitably leads to asymptotic saturation of the TFF in accordance to pQCD and pointed out for the first time that the half-saturated TFF and its asymptotic limit are intimately related. In conjunction with the $1\sigma$ confidence ellipse, we used this new calibration condition on the parameter $B$ to improve the fitting values of $B$ and $C$ with respect to third-party events outside the coverage of the Belle measurements. We argued that Fit(B) can be interpreted as the result of a feedback-loop mechanism [12] which provides asymptotic saturation amounting to a constant. Finally, we worked out a conformity protocol which fares Fit(A) and Fit(B) against selected QCD-based criteria to set key benchmarks on the data-driven exploration of TFF predictions by improving the quality monitoring of forthcoming data and its mathematical modeling.

*Epicrisis of the analysis.* We compared two different representations of pion-photon TFF growth curves: one with inhibition and the other one uninhibited. There are many systems in nature that show such behavior: The Michaelis-Menten equation (Kinetics of enzyme-catalyzed reactions), the Monod equation (The growth of bacterial cultures), etc. We argued that similarly to the above, an inhibited mathematical model (termed Fit(B)) of the Belle data gives rise to asymptotic saturation confirming the QCD prediction. We showed that this saturation phenomenon of the $\pi - \gamma$ TFF resembles a self-stabilized system with a feedback-loop mechanism in the context of control theory. One observes stability of the TFF curve in the sense that each time it is disturbed by a measurement, it has the ability to return to a state close to it by applying the feedback operator (11).

*Acknowledgments.* I thank Dr. Sonja Bastian for discussions on mathematical models of biomedical systems.

---

* Electronic address: `stefanis@tp2.ruhr-uni-bochum.de`

[1] G. P. Lepage and S. J. Brodsky, Phys. Rev. **D22**, 2157 (1980).

[2] S. J. Brodsky and G. P. Lepage, Phys. Rev. **D24**, 1808 (1981).

[3] P. A. Zyla et al. (Particle Data Group), PTEP **2020**, 083C01 (2020).

[4] H. J. Behrend et al. (CELLO), Z. Phys. **C49**, 401 (1991).

[5] J. Gronberg et al. (CLEO), Phys. Rev. **D57**, 33 (1998), hep-ex/9707031.

[6] B. Aubert et al. (BaBar), Phys. Rev. **D80**, 052002 (2009), 0905.4778.

[7] S. Uehara et al. (Belle), Phys. Rev. **D86**, 092007 (2012), 1205.3249.

[8] C. F. Redmer (BESIII), in *13th Conference on the Intersections of Particle and Nuclear Physics (CIPANP 2018) Palm Springs, California, USA, May 29-June 3, 2018* (2018), 1810.00654.

[9] M. Ablikim et al., Chin. Phys. C **44**, 040001 (2020),

1912.05983.
[10] N. G. Stefanis, Phys. Rev. D **102**, 034022 (2020), 2006.10576.
[11] N. G. Stefanis, A. P. Bakulev, S. V. Mikhailov, and A. V. Pimikov, Phys. Rev. **D87**, 094025 (2013), 1202.1781.
[12] N. Wiener, *Cybernetics; or, Control and communication in the animal and the machine* (M.I.T. Press, New York, 1961), ISBN 9780262730099.
[13] V. M. Braun, A. N. Manashov, S. Moch, and J. Schoenleber, Phys. Rev. D **104**, 094007 (2021), 2106.01437.
[14] J. Gao, T. Huber, Y. Ji, and Y.-M. Wang, Phys. Rev. Lett. **128**, 062003 (2022), 2106.01390.
[15] G. P. Lepage and S. J. Brodsky, Phys. Lett. **B87**, 359 (1979).
[16] L. Michaelis and M. L. Menten, Biochemische Zeitschrift **49**, 339 (1913).
[17] L. Michaelis and M. M. L. Menten, FEBS Letters **587**, 2712 (2013).
[18] S. J. Brodsky, F.-G. Cao, and G. F. de Téramond, Phys. Rev. **D84**, 033001 (2011), 1104.3364.
[19] A. P. Bakulev, S. V. Mikhailov, A. V. Pimikov, and N. G. Stefanis, Phys. Rev. **D86**, 031501(R) (2012), 1205.3770.
[20] S. V. Mikhailov, A. V. Pimikov, and N. G. Stefanis, Phys. Rev. D **103**, 096003 (2021), 2101.12661.
[21] A. V. Efremov and A. V. Radyushkin, Theor. Math. Phys. **42**, 97 (1980).
[22] V. L. Chernyak and A. R. Zhitnitsky, Phys. Rept. **112**, 173 (1984).
[23] A. P. Bakulev, S. V. Mikhailov, and N. G. Stefanis, Phys. Lett. **B508**, 279 (2001), [Erratum: Phys. Lett. B590, 309 (2004)], hep-ph/0103119.
[24] S. S. Agaev, V. M. Braun, N. Offen, and F. A. Porkert, Phys. Rev. **D83**, 054020 (2011), 1012.4671.
[25] N. G. Stefanis, Phys. Lett. **B738**, 483 (2014), 1405.0959.
[26] G. S. Bali, V. M. Braun, S. Bürger, M. Göckeler, M. Gruber, F. Hutzler, P. Korcyl, A. Schäfer, A. Sternbeck, and P. Wein (RQCD), JHEP **08**, 065 (2019), [Addendum: JHEP 11, 037 (2020)], 1903.08038.
[27] J. Hua et al. (Lattice Parton), Phys. Rev. Lett. **129**, 132001 (2022), 2201.09173.
[28] X. Gao, A. D. Hanlon, N. Karthik, S. Mukherjee, P. Petreczky, P. Scior, S. Syritsyn, and Y. Zhao, Phys. Rev. D **106**, 074505 (2022), 2206.04084.
[29] L. Chang, I. C. Cloet, J. J. Cobos-Martinez, C. D. Roberts, S. M. Schmidt, and P. C. Tandy, Phys. Rev. Lett. **110**, 132001 (2013), 1301.0324.
[30] V. M. Braun, A. N. Manashov, S. Moch, and M. Strohmaier, JHEP **06**, 037 (2017), 1703.09532.
[31] W. Altmannshofer et al. (Belle-II), PTEP **2019**, 123C01 (2019), [Erratum: PTEP 2020, 029201 (2020)], 1808.10567.
[32] N. G. Stefanis and A. V. Pimikov, Nucl. Phys. **A945**, 248 (2016), 1506.01302.
[33] V. M. Braun and I. E. Filyanov, Z. Phys. **C44**, 157 (1989), [Yad. Fiz. 50, 818 (1989)].
[34] N. G. Stefanis, S. V. Mikhailov, and A. V. Pimikov, Few Body Syst. **56**, 295 (2015), 1411.0528.
[35] B. Melić, D. Müller, and K. Passek-Kumerički, Phys. Rev. **D68**, 014013 (2003), hep-ph/0212346.
[36] S. V. Mikhailov, A. V. Pimikov, and N. G. Stefanis, Phys. Rev. **D93**, 114018 (2016), 1604.06391.
[37] I. I. Balitsky, V. M. Braun, and A. V. Kolesnichenko, Nucl. Phys. **B312**, 509 (1989).
[38] A. Khodjamirian, Eur. Phys. J. **C6**, 477 (1999), hep-ph/9712451.
[39] C. Ayala, S. V. Mikhailov, and N. G. Stefanis, Phys. Rev. D **98**, 096017 (2018), [Erratum: Phys. Rev. D 101, 059901 (2020)], 1806.07790.
[40] S. Mikhailov, A. Pimikov, and N. G. Stefanis, EPJ Web Conf. **258**, 03003 (2022), 2111.12469.
[41] A. Rohatgi, *Webplotdigitizer: Version 4.4* (2020), URL `https://automeris.io/WebPlotDigitizer`.

## SUPPLEMENTAL MATERIAL

### A. QCD theoretical background

Here we present a short exposition of the current status of the QCD calculations related to the perturbative (Subsection A) and nonperturbative (Subsection B) content of the TFF. The QCD-based TFF predictions are discussed in subsection C, while the data analysis related to Fig. 2 is carried out in Subsection D.

*Hard-scattering scheme.* The hard-scattering amplitude $T$ in Eq. (3) contains the short-distance quark-gluon interactions, whereas the large-distance nonperturbative effects are included in $\varphi_\pi(x, \mu_{\rm F}^2)$, which denotes the pion DA taken at the factorization scale $\mu_{\rm F}$. The first quantity is calculable as a power-series expansion in the strong coupling $a_s = \alpha_s(\mu_{\rm R})/4\pi$ in pQCD yielding

$$T = T^{(0)} + a_s T^{(1)} + a_s^2 T^{(2)} + \ldots\,, \tag{12}$$

where the superscript indicates the number of loops and the renormalization scale is set for simplicity equal to the factorization scale: $\mu_{\rm R} = \mu_{\rm F} = \mu$.

The twist-two pion DA can be expressed in terms of the eigenfunctions of the evolution equation at one loop [1, 21]

$$\varphi_\pi^{(\text{tw-2})}(x, \mu^2) = \psi_0(x) + \sum_{n=2,4,\ldots}^{\infty} a_n(\mu^2)\psi_n(x)\,, \tag{13}$$

where $\psi_0(x) = 6x(1-x) \equiv 6x\bar{x}$ is the asymptotic pion DA $\varphi_\pi^{\rm asy}$ and the higher eigenfunctions are defined through the Gegenbauer polynomials $\psi_n(x) = 6x\bar{x}C_n^{(3/2)}(x - \bar{x})$. The coefficients $a_n$ are related to the moments

$$\langle \xi^N \rangle_\pi \equiv \int_0^1 \varphi_\pi^{(\text{tw-2})}(x, \mu^2)(x - \bar{x})^N dx\,, \tag{14}$$

where $\xi = x - \bar{x}$ and $N = 2, 4, \ldots$. They can be determined using different nonperturbative techniques, e.g., QCD sum rules [22–25], lattice QCD calculations, for instance, [26–28], holographic AdS/QCD [18], Dyson-Schwinger equations [29], etc. (see next paragraph). To get the TFF at the experiment momenta, pQCD evolution [1, 21] has to be employed. In the limit $Q^2 \to \infty$ the anomalous dimension is $\gamma_0 = 0$ so that the pion DA evolves to the asymptotic form $\varphi_\pi^{\rm asy}$ and one obtains [2] Eq. (6).

*Higher loops.* Recently, the total two-loop coefficient function $T^{(2)}$ of the leading-twist contribution to the TFF was calculated by two independent groups using different methods but obtaining coinciding analytical results [13, 14]. This establishes the complete knowledge of the coefficient function of the TFF at the NNLO level of pQCD, enabling the inclusion of all radiative corrections up to this order. Combining the full NNLO coefficient function with the three-loop anomalous dimensions, calculated in [30], predictions for the scaled TFF at experiment scales were obtained [13, 14]. The calculated theoretical uncertainties are comparable with the sum in quadrature of the statistical and systematic errors expected from the upcoming Belle II experiment at the SuperKEKB collider. In [31] it is claimed that the measurements for the high $Q^2 > 20$ GeV$^2$ region may be even a factor 3 to 5 times more precise relative to Belle (see Fig. 198 in [31] and related comments).

### B. QCD Nonperturbative input

On the nonperturbative side, one attempts to improve the quality of the pion DA using information from lattice QCD approaches. Until now, only constraints for the second moment of the pion DA with maximally next-to-next-to-leading order (NNLO) accuracy have been computed on the lattice [26]. They favor a coefficient $a_2$ with a central value around $\sim 0.116$ at $\mu_2 = 2$ GeV. However, to determine the shape of the pion DA more reliably, at least its kurtosis is needed. This is defined by means of the fourth moment (i.e., the $a_4$ coefficient) and contains information on the tails of the distribution, rather than its peak [32]. Therefore, the alternative lattice approach based on large-momentum effective theory (LaMET) [27] is welcome because it provides information on the pion DA as a whole, albeit its endpoint behavior still contains rather large uncertainties. Further exploration is needed to extract accurate values of the Gegenbauer moments $a_{n>2}$. In contrast, the central part of the obtained pion DA is more restricted and supports a broad unimodal profile, see also [28] for quite similar results.

It is remarkable that the platykurtic (pk) pion DA [25], which embraces by conception a broad unimodal profile at $x = 1/2$ with suppressed endpoint regions $x = 0, 1$, gives rise to an $\Omega$-shaped distribution function. This DA complies (within errors) rather well with the results of the mentioned lattice approaches, see [10] for details. Besides, the pk DA satisfies at the midpoint $\varphi^{(\text{tw-2})}_{\pi/\text{pk}}(x = 1/2, \mu_1) = 1.264$ the constraint from the LCSR calculation in [33]: $\varphi^{(\text{tw-2})}_{\pi}(x = 1/2, \mu_1) = 1.2 \pm 0.3$ at $\mu_1 = 1$ GeV. For a detailed discussion of the derivation of the platykurtic DA and its remarkable characteristics, we refer to [25, 32, 34].

### C. QCD-based TFF predictions

Partial two-loop contributions at NNLO to the TFF in the $\overline{\text{MS}}$ scheme were obtained before in [35, 36] and were used in [10] to calculate updated TFF predictions within a theoretical scheme which makes use of a state-of-the art implementation of light-cone sum rules (LCSR) [37, 38]. The margin of the total theoretical uncertainty for the bimodal BMS pion DA's [23] and that of the pk DA [25] takes into account the inaccuracy of the pion DA modeling and the only missing two-loop term within this framework. This error margin can be mitigated from below by using the complete NNLO radiative correction from [13, 14] which is positive.

In synergy with a two-loop evolution scheme, which takes into account the crossing of heavy-quark mass thresholds by matching appropriately the flavor number of active flavors in the strong coupling, TFF predictions at the twist-six level were derived within this QCD-based framework for a variety of pion DAs treating them all on equal footing [10]. In general, all considered DA-based predictions agree in trend above 10 GeV$^2$ with the Belle data [7] while being in tension with the rapid growth of the *BABAR* data above this scale [7]—at least at the level of $2\sigma$. Thus, as a result, the extraction of the asymptotic behavior of the TFF from the existing data, still poses for theory an enduring challenge—see [19] for a classification of various theoretical predictions with respect to their high-$Q^2$ behavior relative to the data. It is noteworthy that the TFF calculated with the platykurtic pion DA provides good agreement with most measurements supporting an inhibited large-$Q^2$ behavior [10], while being also in good agreement with the fast growth of the BESIII data at very low $Q^2$ values [20, 39, 40]. The described TFF behavior aligns with the present analysis and the mathematical model in Eq. (4) which incorporates inhibition acting differently as $Q^2$ varies.

### D. Data analysis related to Fig. 2

This figure collects the data and metrics (histograms) of the experiments surveyed in this work.

TABLE III: Main statistics of the TFF central values obtained in the experiments shown in the right panel of Fig. 2.

| Experiment | Maximum\|bin | Median\|bin | Mean\|bin |
|---|---|---|---|
| CELLO [4, 10] | 0.163\|8 | 0.095\|5 | 0.112\|6 |
| CLEO [5] | 0.167\|8 | 0.145\|7 | 0.138\|7 |
| *BABAR* [6] | 0.285\|14 | 0.187\|9 | 0.181\|9 |
| Belle [7] | 0.236\|12 | 0.173\|9 | 0.174\|9 |
| BESIII [8, 20] | 0.164\|8 | 0.115\|6 | 0.108\|5 |

*Center panel.* This panel shows the results of the existing measurements (52 in total) of the TFF $Q^2F^{\gamma\pi}(Q^2)$ vs $Q^2$. They have been obtained in the following single-tag experiments: CELLO [4], (5 points for $Q^2 < 2.2$ GeV$^2$), CLEO [5] (15 points in the range from 1.6 GeV$^2$ to 8.0 GeV$^2$), *BABAR* [6] (17 points between 4 GeV$^2$ and 40 GeV$^2$), and Belle [7] (15 points from 4.0 GeV$^2$ up to 40.0 GeV$^2$). A table with the central values together with the associated error margins can be found in [10], while the main statistical metrics of the considered data sets are collected in Table III. The preliminary BESIII data [8, 9] (18 points in the range $[0.057-1.35]$ GeV$^2$) are also included using the values extracted in [20] using the tool PlotDigitizer [41].

*Top panel.* Graphical representation of the spread of TFF events from different single-tag experiments assembled in 20 bins over the momentum range $Q^2 \in [0.05, 40]$ GeV$^2$ and using a bin width of 2 GeV$^2$. Bins in blue color, extending further to the right, collect the combined counts of measurements at CELLO [4] CLEO [5] *BABAR* [6], and Belle [7]. The blocks containing the data of the BESIII experiment [8] are shown separately in red color to indicate their preliminary status. Note that data points at the intersection of two bins are counted as usual in the next higher bin. One observes that the frequency of events above 10 GeV$^2$ is rather low, less than 35% of the total amount of data with several silent intervals in-between. In fact, more than 50% of the measurements were performed below [10-12] GeV$^2$ while the $Q^2$ intervals above 15 GeV$^2$ are only scarcely populated with data bearing rather large errors. The positive skewness of this data distribution signifies the extent of asymmetry between low and large $Q^2$ values emphasizing the need for more dense measurements above 10 GeV$^2$.

*Right panel.* This panel shows histograms in blue color displaying the frequency counts of measured values of $Q^2F^{\gamma\pi}(Q^2)$ from [4], [5], [6], and [7]. The preliminary data from [8] are included in red color on top of the others. The distribution of the measurements covers the dynamical range $[0.01-0.31]$ GeV and is sampled in 15 bins from bottom to top along the vertical axis using an increment of 0.02 GeV. The horizontal axis shows the frequency values of TFF events within each bin.

The main statistical measures are given in Table III. The key observation is that most measurements are clustered around a common mode in the interval $Q^2F(Q^2) \in [0.15-0.17]$ GeV (bin 8), though no statistical combination of independent data has been involved. This mode value takes into account a total number of 16(17) events from CELLO (1), CLEO (4), *BABAR* (6), Belle (5), and BESIII (1) and represents the largest share at $17/52 \approx 33\%$. Remarkably, both measurements (Belle and *BABAR*) yield very close mean values within bin 9, see Table III, and are not dominated by their high-end behavior in bins 12 and 14, respectively. Doubling the bin size, one finds that the interval $[0.15-0.19]$ GeV constitutes the highest percentage of the measured TFF values at 50%. This estimate matches the possibility of saturating behavior of the TFF starting in the momentum range $[10-22]$ GeV$^2$. We verified that these observations are not sensitive to a rebinning of the data.